\RequirePackage{fix-cm}
\documentclass[smallextended]{svjour3}       
\smartqed  
\usepackage{graphicx}
\usepackage{multirow}
\usepackage{multicol} 
\usepackage{amssymb}
\usepackage{amsmath}
\usepackage{mathtools}
\usepackage{algorithmic}
\usepackage{url}
\usepackage{graphicx}
\usepackage{hyperref}
\usepackage{color}
\usepackage[top=2cm, bottom=2cm, left=2cm, right=2cm]{geometry}
\usepackage[ruled,vlined]{algorithm2e}
\usepackage{array}
\usepackage{comment}
\usepackage{slashbox}
\usepackage{natbib}

\newcolumntype{H}{>{\setbox0=\hbox\bgroup}c<{\egroup}@{}}

\usepackage{etoolbox}
\apptocmd{\thebibliography}{\renewcommand{\sc}{}}{}{}
\begin{document}

\title{Transfer of codebook latent factors for cross-domain recommendation with non-overlapping data}


\author{Sowmini Devi Veeramachaneni  \and Arun K Pujari \and Vineet Padmanabhan  \and Vikas Kumar}

\institute{Sowmini Devi Veeramachaneni, Arun K Pujari \at
              \`Ecole Centrale School of Engineering, Mahindra University, Hyderabad, Telangana, India \\
              Tel.: +91-8985061848\\
              \email{sowmiveeramachaneni@gmail.com, arun.k.pujari@gmail.com}           
           \and
           Vineet Padmanabhan \at
               School of Computer and Information Sciences, University of Hyderabad, Hyderabad, Telangana, India\\
              \email{vineetcs@uohyd.ernet.in}
              \and
              Vikas Kumar \at
               University of Delhi, Delhi, India\\
               \email{vikas007bca@gmail.com}
}

\maketitle

\begin{abstract}
Recommender systems based on \emph{collaborative filtering} play a vital role in many E-commerce applications as they guide the \emph{user} in finding their \emph{items} of interest based on the user's past transactions and feedback of other similar customers. \emph{Data Sparsity} is one of the major drawbacks with collaborative filtering technique arising due to the less number of transactions and feedback data. In order to reduce the sparsity problem, techniques called transfer learning/cross-domain recommendation has emerged. In transfer learning methods, the data from other dense domain(s) (source) is considered in order to predict the missing ratings in the sparse domain (target). In this paper, we come up with a novel transfer learning approach for cross-domain recommendation, wherein the cluster-level rating pattern(codebook) of the source domain is obtained via a co-clustering technique. Thereafter we apply the Maximum Margin Matrix factorization(MMMF) technique on the codebook in order to learn the user and item latent features of codebook. Prediction of the target rating matrix is achieved by introducing these latent features in a novel way into the optimisation function. In the experiments we demonstrate that our model improves the prediction accuracy of the target matrix on benchmark datasets.


\end{abstract}

\section{Introduction}
\vspace{-0.2cm}
The key idea of a recommender system (RS) \cite{BOBADILLA2013109, 10.5555/1941884, 10.5555/2931100} is to provide useful information to users regarding the products/items that would be interested in. Collaborative filtering (CF) is one of the common techniques used in recommendation engines to learn user profiles so that preferred items can be recommended. In CF, a recommendation is made for a user (usually called the target user) by taking into consideration the known preferences of other users who are much alike to the target user.  
Matrix factorization~\cite{koren2009matrix,wu2007collaborative} is widely considered as one of the most promising collaborative filtering techniques.

 MF finds the user and item latent features from a given user-item rating matrix, and the product of these latent feature vectors yields the approximation/prediction of the rating matrix.
Let us suppose that we are given $X \in \mathbb{R}^{m_1 \times n_1}$ which is a user-item rating matrix. The number of users is denoted by $m_1$ and $n_1$ denotes the items. The idea in MF is to find two matrices, $U \in \mathbb{R}^{m_1 \times l}$ and $V \in \mathbb{R}^{n_1 \times l}$ ($l$ is the number of latent factors), such that $U \times V^T = \hat{X} \approx X$ (i.e., the product is approximately equal to $X$. ), on observed ratings ($\mathcal{O}$). %
The problem can be formulated as follows,
\begin{equation*}
 Minimize ~~~ \mathcal J = \sum\limits_{(i,j)\epsilon\mathcal O} Loss(x_{ij}, u_iv_j)
\end{equation*}
Here, $Loss(\cdot)$ is the loss fuction that estimates the difference between the actual rating ($x_{ij}$) and the predicted rating ($\hat{x_{ij}}$). In several factorization models we can see that the loss function being considered is the \emph{squared error}.
In collaborative filtering, when the ratings are discrete \{1, 2,...,r\}, Maximum Margin Matrix Factorization (MMMF)~\cite{srebro2004maximum, 6973968, 10.1007/978-3-319-42911-3_14} is shown to be successful. MMMF techniques uses \emph{hinge loss} as the loss function and this helps in constraining the trace norms of $U$ and $V$  rather than constraining the dimensions, through regularisation.
The objective is to discover the latent factor matrices of users ($U\in \mathbb{R}^{m_1\times l}$) and items ($V\in \mathbb{R}^{n_1\times l}$), and $r-1$ thresholds $\theta_{iq}$ ($1\leq q\leq r-1$) for each user $i$ by minimizing the following optimization function.
\begin{equation}\label{mmmf}
\mathcal{J}(U,V,\Theta) = \sum_{(i,j) \in \mathcal{O}}\sum_{q=1}^{r-1}h(\mathcal{T}_{ij}^q(\theta_{iq} - u_iv_j^T)) + \lambda(||U||_F^2+||V||_F^2)
\end{equation}

where,  $\lambda >0$ is regularization parameter, 
\begin{equation*}
    \mathcal{T}_{ij}^q = \begin{cases}
-1 & \text{if $q < x_{ij}$}\\
+1 & \text{if $q \geq x_{ij}$\hspace{4.15cm}}
\end{cases}
\end{equation*}
and $h(\cdot)$ is a smooth hinge-loss function with the following definition:\\
\begin{equation}\label{hinge}
h(f) = \begin{cases}
0 & \text{if $f$ $\geq$ 1} \\
\frac{1}{2}(1 - f)^{2} & \text{if $0 < f < 1$}\\
\frac{1}{2} - f & \text{otherwise.\hspace{3.2cm}}
\end{cases}
\end{equation}
The optimisation function for MMMF as given in the  equation (Eq. \ref{mmmf}) 
does not require $u_iv_j^T$ and $x_{ij}$ to be closer. It expects $u_iv_j^T$ to be as small as possible if $q \geq x_{ij}$ and as large as possible if $q < x_{ij}$, when compared with $\theta_{iq}$. One can solve the optimization function given in Eq. (\ref{mmmf}) using gradient descent. 

Though MMMF based collaborative filtering techniques perform well with discrete data it is confined to a single domain and fails to account for user-item interaction when multiple domains are involved. They also do not perform well when the number of ratings are low resulting in \emph{data sparsity} problem. Transfer learning methods~\cite{pan2010survey} have been developed to address these concerns.

Transfer learning helps in building a predictive model by extracting and transferring common knowledge when multiple domains are involved. These domains are usually referred in the literature as \emph{source} and \emph{target}.
When transfer learning techniques are applied in the domain of recommender systems, two critical problems needs to be addressed for successful knowledge transfer. 
These are 1) how to account for knowledge transfer if both the domains have shared users/items and 2) accounting for transferring of knowledge when domains do not have shared users/items. As far as the first problem is concerned, it is quite difficult for such a scenario to exist in real world as the users/items in one system may not be present in another. Even otherwise, if such correspondence exists, mapping becomes a difficulty  as the users/items may have different names in different systems. 
The second problem is pretty hard to address and in this study we employ a representative approach to do so.

In this paper, we come up with a novel approach of transferring the learnt knowledge from \emph{source} to the \emph{target} by assuming that the source and target domains have some implicit correlation. The learnt knowledge of source which are the latent factors of codebook get transferred to the target via hinge loss. Experimental results demonstrate the superiority of the proposed approach for cross-domain collaborative filtering as compared to other major methods in transfer learning which are codebook-based. 

The remainder of the paper is structured as follows: In Section \ref{TL}, we talk about related work. The proposed strategy is outlined in Section \ref{prop}. Section \ref{expt} discusses the experimental results. Section \ref{conclusion} contains the conclusion and future work.
\section{Related Work}\label{TL}
Though collaborative filtering based recommender systems have become the norm these days, they have trouble in making accurate recommendations due to data sparsity problem, i.e., very little actual information available. To address the data sparsity issue in recommender systems, \emph{transfer learning}~\cite{pan2010survey, pan2016survey, zhao2013active} techniques (for cross domain recommender systems) has been proposed in the literature. As mentioned earlier, in transfer learning there is a source domain which is usually considered as a \emph{dense} domain from which knowledge is transferred to the target domain which is usually \emph{sparse}.  
For example, suppose that a particular user has watched lots of movies and have rated many of the movies he has seen. Suppose also that the same user has read many books and has given very few ratings in the books domain  but at the same time wants a book that is of interest to him to be recommended. In such a  scenario, the idea is to make use of the users' ratings in the movie domain such that a particular book of his/her interest can be recommended.

One of the major issues involved in developing transfer learning techniques for recommendation purpose is to establish a bridge between the domains that are involved, so that knowledge can be transferred from \emph{source} domain to the \emph{target} domain. Domains can be linked and the transfer can happen explicitly via inter-domain similarities, common item attributes, etc. The transfer can happen also implicitly via shared user latent features or item latent features or by rating patterns which can be transferred between the domains. In~\cite{Chung:2007:IPR:1282100.1282113}, a framework was proposed where the items that are relevant in the source domain are picked based on the common attributes they have with the target domain (user-interested domain). In this way, the inter-domain links were built utilising the common item attributes, however no overlap of users/items was required between the domains. 
On the other side the transfer of knowledge through shared latent features (of users/items) is addressed in~\cite{pan2010transfer}. The idea proposed in this paper\cite{pan2010transfer} is to learn the hidden features present in the users and items of the source domain so that they can be integrated into the target rating matrix during the factorization process via regularisation. 
The success of this procedure is dependent on the existence of common users or items. 
In~\cite{Pan:2011:TLP:2283696.2283784}, the latent properties of source and target are shared in a collective way. Here, rather than learning the latent features from source and utilizing them in the target, a technique is proposed wherein the latent features are simultaneously learnt from both the domains. A method called matrix tri-factorization is used to construct the shared latent space with the condition that from both the domains the users and items needs to be identical.

There are other set of methods in which \emph{rating patterns} are analysed and transferred rather than latent features. These methods can be used in scenarios wherein users/items are not common between the domains.
Rating patterns stem from the assumption that among the ratings of groups of users and groups of items a correlation could exist.   
One such method is codebook transfer (CBT)~\cite{li2009can}, where the main assumption is that, though the users/items are different across systems, the clusters (groups - based on age, interest,...) of them behave similarly. It is an adaptive method which consists of mainly two steps. One is the codebook construction and the second step is filling the target matrix by transferring the learnt codebook. As part of the initial step, the users and items that belong to the dense source domain are co-clustered \cite{Ding:2006} to get the cluster-level rating pattern(rating pattern at the cluster level). This pattern is called the \emph{codebook} which consists of the mean rating of each of the co-clusters of users and items.
Following that, the codebook is transferred to the target domain by expanding the values of the codebook. To do the same, users and items of the target domain needs to be mapped (to co-clusters) 
and this can be done by minimizing the quadratic loss which can be expressed as,
\begin{equation}\label{trifact2}
\min_{F_{1}\in\{0,1\}^{m_1\times k_1}, F_{2}\in\{0,1\}^{n_1\times k_2}}||[X-F_{1}CF_{2}^T]\odot W||_{F}^{2}\quad \text{s.t.,$F_1\bf{1}$ = $1$, $F_2\bf{1}$ = $1$.}
\end{equation}

Here, $X_{m_1\times n_1}$ represents the user-item rating matrix of target domain.
$C_{k_1\times k_2}$ stands for codebook (rating pattern at cluster-level) which is learned from the source domain. The codebook is fixed and used to learn the cluster membership matrices of users (${F_1}_{m_1\times k_1}$) and items (${F_2}_{n_1\times k_2}$) of target data. 
A value of $1$ in the indicator matrix $W$ of size $m_1\times n_1$ shows the existence of the rating in the original rating matrix whereas a value of $0$ indicates the absence of the rating. 
The idea of code book transfer is to discover a common latent space wherein the knowledge gained in the form of $C$ (source domain data) can be used to enhance the recommendation in the target domain. i.e., here the ratings are transferred in the condensed form (codebook). 

 In~\cite{Li:2009:TLC:1553374.1553454}, authors have proposed a method named rating-matrix generative model which uses a probabilistic framework and fill the missing
ratings of target domain by considering the rating data from multiple source rating matrices to construct the rating pattern. \cite{Moreno:2012:TTL:2396761.2396817} extends the CBT by considering multiple source domains and checking different combinations of user/item clusters. It creates seperate codebooks for every source domain and extracts the relatedness between the target and each of the sources. It is based on the linear mixture of different codebooks in which the learning of weights is done by the minimization of target domain prediction error. A relaxation related to the assumption of a fully dense source domain rating matrix is taken into consideration  in~\cite{8233662, He:2018:RTL:3159652.3159675}. A different way of generating the codebook has been outlined in~\cite{ji2016improving, DBLP:journals/asc/VeeramachaneniP19, VEERAMACHANENI2022102002}. In these methods, by making use of the technique of matrix factorisation the user and item latent factors are generated from the source domain. The user and item latent factors thus generated are used to obtain the user latent facor group and item latent factor group. Codebook is generated by multiplying the mean latent vectors of the group. 

The methods as outlined in~\cite{itcf, PAN201684} takes care of scenarios wherein the feedback data of target and source are of different types. The users and items in the target and auxiliary(source) domains are assumed to be the same in~\cite{itcf}. 
Two sets of source data are taken into consideration in~\cite{tbt}, one of which shares a common set of users with the target data and the other of which shares a common set of items with the target data. The source data's latent factors are extracted, and similarity graphs are constructed from these latent factors. Both latent factors and similairty graphs are then transferred to the target data.

\section{Proposed Approach}\label{prop}
\begin{figure*}
\centering
\includegraphics[scale=0.61]{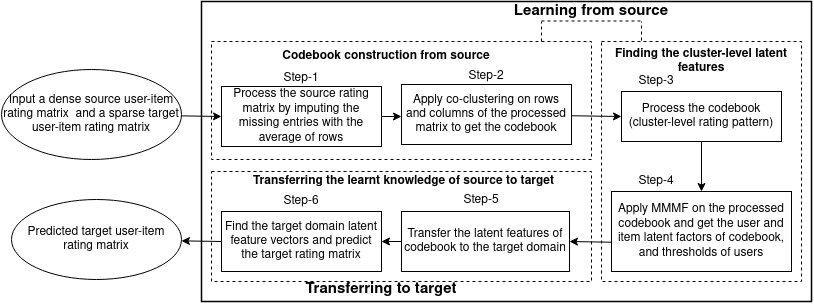}
\caption{Illustration of the proposed method}
\label{blk_diag}
\end{figure*}

Let there be two user-item rating matrices of different domains say $X_{m_1\times n_1}$ (target domain matrix), $Y_{m_2\times n_2}$ (source domain matrix). Here $m_1$, $m_2$ is the number of users and $n_1$, $n_2$ is the number of items, and the entries of the matrices are the ratings given by the users to the items. Our goal is to predict the missing entries of the target domain more accurately using the source domain data. Figure-\ref{blk_diag} gives the sequential steps of the proposed method. Initially, we fill the missing entries in the source rating matrix with the mean of the ratings of that row (Step-1) and denote the filled-in rating matrix as $Y'$. In Step-2, we apply the co-clustering on the filled-in rating matrix ($Y'$) in order to get the rating pattern at the cluster-level called as codebook ($C_{k_1\times k_2}$). Once the codebook is obtained, we process the codebook (Step-3) by removing some of the entries of codebook and replacing by $0$. Processing of codebook is done by comparing it with filled-in rating matrix as follows. Take the entry of the codebook which indicates the average of ratings given by a cluster of the users to some group of items. Compare the value with the entries of the particular users (forming a cluster) and particular items (of the cluster) of the filled-in rating matrix. If the number of entries containing the same value is more than some specific  threshold percentage ($th$) then keep it, else we remove and replace it as zero. As there are real values in codebook and filled-in rating matrix, we don't check for the values to be exact, but instead we check for their difference to be small. The difference (error) is compared using some margin $\epsilon$. Calculate the difference between the entries of filled-in matrix and that of codebook, and if more than some threshold percentage ($th$) of entries contain the margin less than or equal to $|\epsilon|$, then keep the entry as it is, else remove the entry. By following this removal of entries we get a partial codebook ($C_p$). Now, apply MMMF (\ref{mmmf}) on the processed codebook ($C_p$) to get the codebook's latent feature vectors of users ($U_{c_{k_1\times l'}}$), and items ($V_{c_{k_2\times l'}}$) alongside a threshold matrix of users ($\Theta_{c_{k_1\times r-1}}$), which is shown in Step-4. 
Once thedr latent features are obtained, we transfer the same to the target domain which is given at Step-5, by minimizing the optimization function given in Eq. \ref{latent}. \textit{As far as we know, there is no previous research which addresses the transfer of latent features of codebook, and also consider hinge loss (Eq. \ref{latent}) as the loss function while transferring the learnt knowledge of the source domain to the target domain.} Our assumption in this work is that there could exist between the source and target domain some implicit correspondence through the user/item latent features ($U_c$, $V_c$) of codebook.
\begin{equation}\label{latent}
\begin{aligned}
\mathcal{J}(\alpha, \beta) ={} & \sum_{(i,j) \in \mathcal{O}}\sum_{q=1}^{r-1}h(\mathcal{T}_{ij}^q(\alpha_i\theta_{\cdot q} - (\alpha_iU_c)(\beta_jV_c)^T)) + \lambda_1(\sum_{i=1}^{m_1} l_1(\sum_{k=1}^{k_1}\alpha_{ik})+ \sum_{j=1}^{n_1}l_1(\sum_{k=1}^{k_2}\beta_{jk})) \\
      & + \lambda_2(\sum_{i=1}^{m_1}\sum_{k=1}^{k_1}l_2(\alpha_{ik})+\sum_{j=1}^{n_1}\sum_{k=1}^{k_2}l_2(\beta_{jk}))
\end{aligned}
\end{equation}
where,\\
  \vspace{-0.3cm}
\hspace{4.2cm}$\mathcal{T}_{ij}^q = \begin{cases}
+1 & \text{if $q \geq x_{ij}$}\\
-1 & \text{if $q < x_{ij}$}
\end{cases}$\\

\vspace{0.8cm}
\hspace{3.5cm}$ l_1(d) = \begin{cases} 
1 - d, & \text{if $d<1$}\\
d-1, & \text{if $d>1$}\\ 
 0, & \text{otherwise}
\end{cases}$\\

\vspace{0.8cm}
\hspace{3.5cm}$ l_2(d) = \begin{cases} 
d, & \text{if $d<0$}\\
 0, & \text{otherwise}
\end{cases}$\\
 $\lambda_1, \lambda_2 >0$ are regularization parameters. $h(\cdot)$ is the smooth-hinge loss defined as given in Eq.(\ref{hinge}). For a given matrix $A$, $A_{\cdot a}$ is the $a^{th}$ column of $A$. $l_1(d)$ ensures the row sum of $\alpha$ and column sum of $\beta$ to be $1$, whereas $l_2(d)$ ensures all elements of $\alpha$ and $\beta$ to be positive.
We have used Gradient Descent technique to optimize the Eq.(\ref{latent}), by updating the variables $\alpha$ and $\beta$. Initially $\alpha$ and $\beta$ are randomly assigned, and then by calculating the gradients of Eq.(\ref{latent}) w.r.t. to $\alpha$ and $\beta$, we update $\alpha$ and $\beta$. By using the updated values of $\alpha$ and $\beta$, the value of the Eq.(\ref{latent}) decreases monotonically and converges to a local minimum. 

The gradients of Eq.(\ref{latent}) w.r.t. variables $\alpha$ and $\beta$ are as follows,

\begin{equation}
\frac{\partial\mathcal{J}}{\partial{\alpha_{ik}}} = \lambda_1l_1^{'}(\sum_{k=1}^{k_1}\alpha_{ik})+\lambda_2l_2^{'}(\alpha_{ik}) + \sum_{q=1}^{r-1}\sum_{(i,j) \in \mathcal{O}}\mathcal{T}_{ij}^q.h'(\mathcal{T}_{ij}^q(\alpha_i\theta_{\cdot q} - (\alpha_iU_c)(\beta_jV_c)^T))(\theta_{kq} - IU_c(\beta_jV_c)^T)
\end{equation}
\begin{equation}
\frac{\partial\mathcal{J}}{\partial{\beta_{jk}}} = \lambda_1l_1^{'}(\sum_{k=1}^{k_2}\beta_{jk})+\lambda_2l_2^{'}(\beta_{jk}) - \sum_{q=1}^{r-1}\sum_{(i,j) \in \mathcal{O}}\mathcal{T}_{ij}^q.h'(\mathcal{T}_{ij}^q(\alpha_i\theta_{\cdot q} - (\alpha_iU_c)(\beta_jV_c)^T))(\alpha_iU_c)(V_{c_k})^{T}
\end{equation}
where,
\begin{equation*}
h'(f) = \begin{cases}
0 & \text{if $f$ $\geq$ 1} \\
f-1 & \text{if $0 < f < 1$}\\
-1 & \text{otherwise.}
\end{cases}
\end{equation*}
\begin{alignat*}{2}
    & \begin{aligned} & l_1^{'}(d) = \begin{cases}
 -1 , & \text{if $d<1$}\\
  1, & \text{if $d>1$}\\ 
 0, & \text{otherwise}
  \end{cases}\\
  \MoveEqLeft[-1]
  \end{aligned}
    & \hskip 6em &
  \begin{aligned}
  &l_2^{'}(d) =  \begin{cases}
 1, & \text{if $d<0$}\\
 0, & \text{otherwise}
  \end{cases} \\[3.3ex]
  \MoveEqLeft[-1]
  \end{aligned}
\end{alignat*}
$I$ is the row vector of dimension $1\times k_1$ containing all $1's$. \\
The updation of $\alpha$ and $\beta$ variables is done as,

\begin{equation*}
\alpha_{ik}^{t+1} = \alpha_{ik}^{t}-a\frac{\partial\mathcal{J}}{\partial{\alpha_{ik}^t}}
\end{equation*}
\begin{equation*}
\beta_{jk}^{t+1} = \beta_{jk}^{t}-a\frac{\partial\mathcal{J}}{\partial{\beta_{jk}^t}}
\end{equation*}
 Here $a$ is the step size. Once we get the converged values of $\alpha$ and $\beta$, we construct the predicted target rating matrix (Step-6) using Equation (\ref{predict}), and map the resultant matrix with the threshold matrix ($\alpha.\Theta_c$) to get the target predicted rating matrix.
\begin{equation}\label{predict}
\hat{X}=W\odot X + [1-W]\odot [\alpha U_c V_c^T\beta^T],
\end{equation}
%
\begin{algorithm}[h]
\caption{Finding latent features of codebook and transferring to target} 
\label{mfclust}
\begin{algorithmic}[1]
\STATE \textbf{Input:}\hspace{10pt} A $ m_2 \times n_2$ source rating matrix $Y$ and a $m_1 \times n_1$ sparse target rating matrix  $X$ \\with $x_{ij}$ known for  $(i,j)\in\mathcal{O}$
\STATE \textbf{Output:}\hspace{2pt} $x_{ij}$ for $(i,j)\notin\mathcal{O}$ 
\STATE Fill the missing entries of each row of $Y$ with the average of the rows and call it as $Y'$
\STATE Apply co-clustering on $Y'$ to get the codebook ($C$)
\STATE Process the codebook to get the partial codebook ($C_p$)
\STATE Find $U_c$, $V_c$, $\Theta_c$ using MMMF (by minimizing Eq. \ref{mmmf}) on $C_p$.
\STATE Use $U_c$, $V_c$, and find $\alpha$, $\beta$ of target domain by minimizing equation (\ref{latent}).
\STATE Using these $\alpha$ and $\beta$, calculate Eq. \ref{predict} and map with the $\alpha\Theta_c$ in order to get the discrete predicted rating matrix ($\hat{X}$).
\vspace{-0.1cm}
\end{algorithmic}
\end{algorithm}
where $\hat{X}$ is the predicted (approximated) target rating matrix. A value of 1 in the indicator matrix $W$ of size $m_1\times n_1$ shows the existence of the rating in the original rating matrix, whereas a value of $0$ indicates the absence of the rating.  
Error calculation for only the observed ratings is ensured through $W$ and the Hadamard product (element wise product) is denoted using $\odot$.
By using the gradient descent technique as given in Eq.(\ref{latent}) a minimal solution can be obtained by updaing  $\alpha$ and $\beta$.
Initially $\alpha$ and $\beta$ are randomly assigned, and then by calculating the gradients of Eq.(\ref{latent}) w.r.t. to $\alpha$ and $\beta$, we update $\alpha$ and $\beta$. By using the updated values of $\alpha$ and $\beta$, the value of the Eq.(\ref{latent}) decreases monotonically until a local minima is reached. Once we get $\alpha$ and $\beta$ by working out with the optimization function (\ref{latent}), we generate the predicted target rating matrix (Step-6) by making use of Equation (\ref{predict}), and map the resultant matrix with the threshold matrix ($\alpha.\Theta_c$) to get the target predicted rating matrix.

To be brief of Figure-\ref{blk_diag}, it depicts the flow of the proposed method, in which \textit{learning from source} and \textit{transferring to target} are the main steps. In the learning stage, the codebook's latent features of source domain are learnt, and in the transferring stage, the learnt knowledge (latent features of codebook) get transferred to the target domain in order to predict the missing ratings of target domain more accurately.
\section{Experimental Analysis}\label{expt}
\texttt{MovieLens-1M}\footnote{https://grouplens.org/datasets/movielens/} is used 
as the \emph{source} dataset and
\texttt{Goodbooks}\footnote{https://github.com/zygmuntz/goodbooks-10k}
is used as the \emph{target} dataset in our experiments. 
We have taken the first 5000 users and 3000 items from the Goodbooks data.
The values of the datasets are in \{0,1,2,3,4,5\}. The value $0$ indicates that the rating is missing, and $1$ indicates the least rating, and $5$ is the highest rating. Table~\ref{dataset} gives the statistics of the datasets. In our experiments, we have divided the data into training (80\%) and testing (20\%) sets. 
\begin{table}[ht]
\centering
\caption{Datasets statistics}
\vspace{0.25cm}
\label{dataset}
\resizebox{10.5cm}{!}{%
\begin{tabular}{|c|c|c|c|}
\hline
\textbf{Dataset} & \textbf{\# of Users} & \textbf{\# of Items} & \textbf{\% of Observed entries} \\ \hline
MovieLens 1M    & 6040 & 3952 & 3.77  \\ \hline
Goodbooks 	& 5000 & 3000 & 1.08  \\ \hline
\end{tabular}
}
\end{table}
\subsection{Evaluation Metrics}
From the literature it can be seen that a variety of collaborative filtering algorithms have been put forward in the last decade or so. The accuracy with which these algorithms can predict a new item/set of items vary. It is often the case that performance evaluation of these collaborative filtering algorithms is based on prediction accuracy. The two most often used measures for computing the prediction accuracy are Mean Absolute Error (MAE) (Eq. (\ref{mae}) and Root Mean Square Error (RMSE) (Eq. (\ref{rmse})). These metrics are based on the difference between the true ratings and the predicted ratings, and it is natural that better performance equals smaller values of RMSE and MAE metrics. 

\begin{equation}\label{rmse}
RMSE = \sqrt{\sum\limits_{(i,j)\epsilon \mathcal O}\frac{{(x_{ij}-\hat x_{ij})}^2}{|\mathcal O|}}
\end{equation}
\begin{equation}\label{mae}
MAE = {\sum\limits_{(i,j)\epsilon \mathcal O}\frac{|(x_{ij}-\hat x_{ij})|}{|\mathcal O|}}
\end{equation}

where $x_{ij}$ is the original rating, $\hat x_{ij}$ is the predicted rating, and $|\mathcal{O}|$ is the number of test ratings.
\subsection{The different Methods used for Comparison}
Some of the baseline methods we use for evaluating the performance of our proposed method can be outlined as follows:

\begin{itemize}
\item \textbf{MMMF~\cite{srebro2004maximum,Jas05}:} Maximum Margin Matrix Factorization (MMMF) is the dominant factorization technique used in collaborative filtering. 
 MMMF is usually applied on the input rating matrix consisting of the user-item ratings. The idea is to find the user and item latent-factor vectors which are of low rank by making use of the existing ratings. 
MMMF can be applied on a single domain only, and hence in our experiments, we applied it on the target domain directly.
\item \textbf{MINDTL \cite{8233662}:} 
In MINDTL, codebook is constructed by taking into consideration the data from all the incomplete source domains. Here codebook for each domain is constructed. 
Following that, the constructed codebooks are linearly integrated and transferred to target, and the missing(absent) values of the target rating matrix gets predicted. As far as our experimental setup is concerned only a single domain is taken into consideration. 

\item \textbf{TRACER \cite{ZHUANG2018287}:} 
In TRACER, data from multiple domains are accounted for and based on this, ratings (which includes missing ratings) for all the source matrices are predicted. Thereafter the predicted knowledge is utilized by transferring it into the target domain. By making use of consensus regularisation during the knowledge transfer process, all the predicted values are forced to be similar. In a way it can be said that in TRACER at the same time learning and transferring happens. 
We have considered a single domain in our experiments and therefore there is no need for consensus regularisation. 

\item \textbf{CBT \cite{li2009can}}: In this approach, the dense part of the source user-item rating matrix is considered, and the missing values of the rows of the dense matrix get imputed using the average of the ratings of particular row (user). The codebook is obtained from the dense user-item matrix by applying the technique of co-clustering. 

In our experiments, unlike in \cite{li2009can}, which consider only the dense part of the input data, the codebook is constructed by making use of the whole source data. Transferring of the learned codebook to the target domain is achieved by  minimizing Eq. (\ref{trifact2}).

\end{itemize}


\begin{figure}[h]
\centering
\includegraphics[scale=1.35]{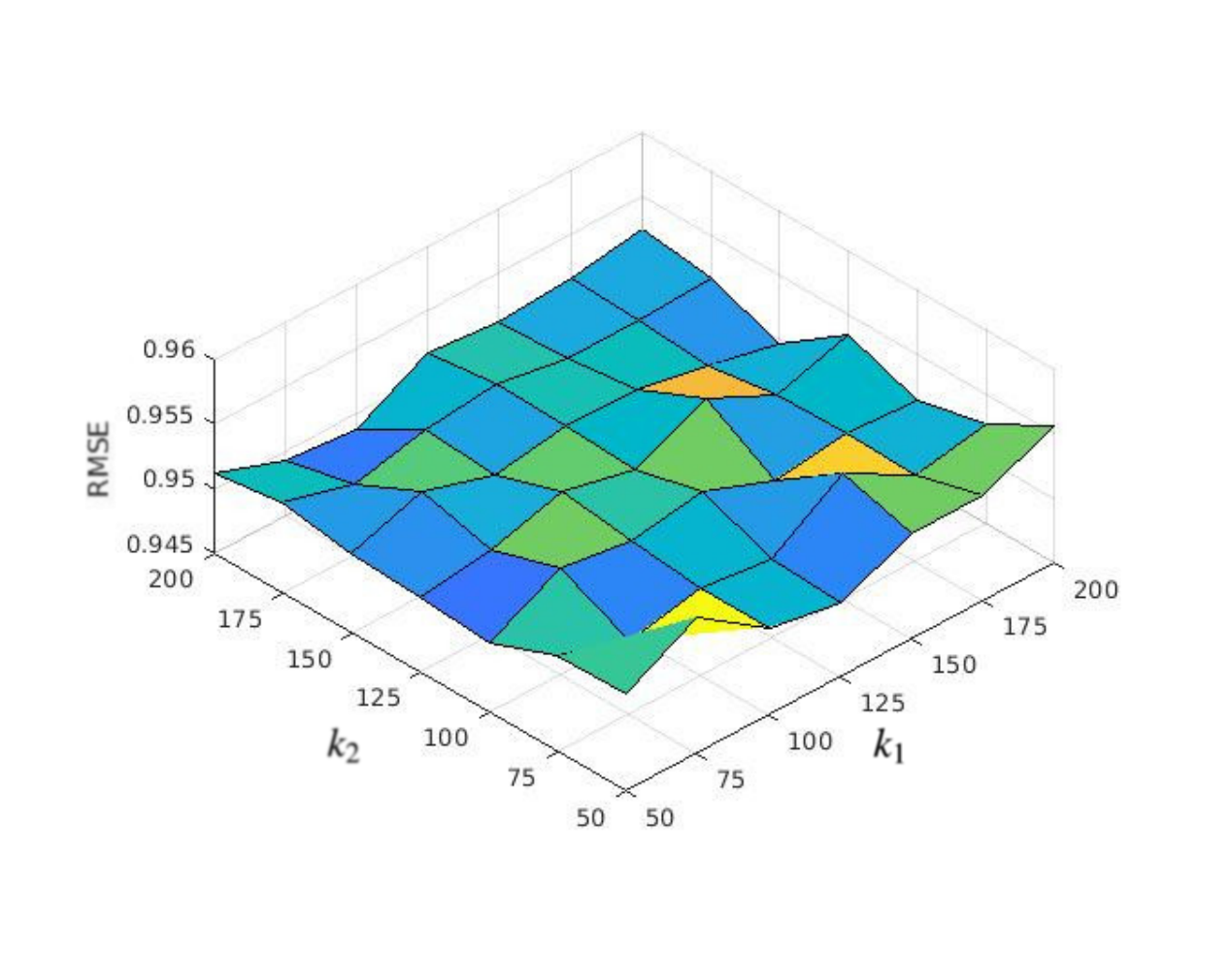}
\vspace{-1.2cm}
\caption{Impact of number of clusters on RMSE of Goodbooks data when MovieLens-1M is considered as source}
\label{fig:rmse}
\end{figure}
\begin{figure}[h]
\centering
\includegraphics[scale=1.35]{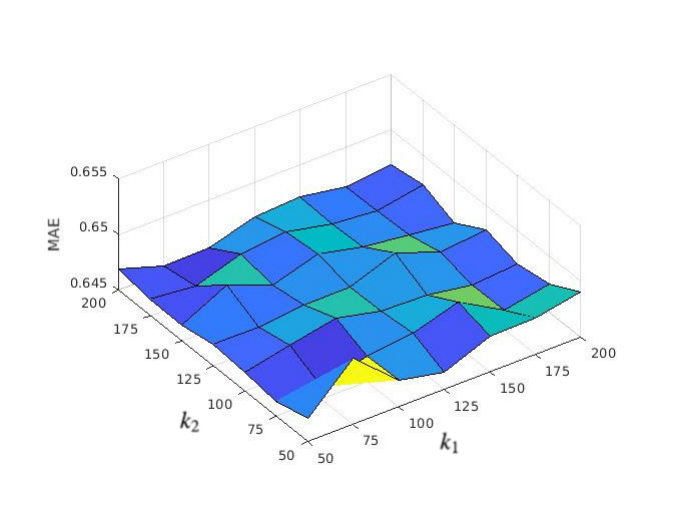}
\vspace{-1.2cm}
\caption{Impact of number of clusters on MAE of Goodbooks data when MovieLens-1M is considered as source}
\label{fig:mae}
\end{figure}


\begin{table}[ht]
\centering
\caption{Values of root mean square error and mean absolute error of baseline methods and our method on Goodbooks data using MovieLens-1M as source}
\vspace{0.25cm}
\label{rmse_mae_prop}
\resizebox{11.5cm}{!}{%
\begin{tabular}{|c|c|c|c|c|c|}
\hline
\backslashbox{Metric}{Method} & MMMF & MINDTL & TRACER & CBT & Proposed \\ \hline
RMSE  & 0.9582 & 1.2794 &  0.9637  & 0.9641 & \textbf{0.9507}   \\ \hline
MAE     & 0.6501  & 0.9232  & 0.7781 & 0.7890 & \textbf{0.6466} \\ \hline
\end{tabular}
}
\end{table}

We have conducted the experiments on MovieLens-1M data (source) and Goodbooks data (target), with varying number of clusters ($k_1$, $k_2$ - 25, 50, 75, 100, 125, 150, 175, 200). Fig. (\ref{fig:rmse}) gives the impact of number of clusters on RMSE and Fig. (\ref{fig:mae}) shows the impact of number of clusters on MAE. Although there is no much change in the metric values with varying number of clusters, in our experiments we have fixed $k_1$ to $150$, $k_2$ to $100$, for which the best performance is achieved.
By fixing the number of clusters, we have also experimented our algorithm by varying the values of threshold ($th$) and margin ($\epsilon$). The range of $th$ (in \%) fall in \{40, 50, 60, 70, 80\}, and the values of $\epsilon$ considered are \{0.1, 0.2, 0.3, 0.4, 0.5\}. The performance of our algorithm is satisfying when the value of $th$ is $50$, and that of $\epsilon$ is $0.2$. Hence, in the experiments of the proposed method, when Goodbooks data is the target and MovieLens-M is source, we have fixed the values of parameters as follows: $k_1$ = 150, $k_2$ = 100, $th$ = 80, $\epsilon$ = $0.3$.

Table~\ref{rmse_mae_prop} shows the RMSE and MAE values on \texttt{Goodbooks} (target) data of baseline methods considered and the proposed method, by considering MovieLens-1M as source. The values reported are the average of five runs.

\section{Conclusions and Future work}\label{conclusion}
We have proposed a novel model for cross-domain recommendation that takes into account the latent features of the source domain's codebook and utilizing them in the target domain, when the domains do not share common users or common items. As the first step, we imputed the missing entries of the source rating matrix with the average of the rows to get the filled-in rating matrix. We made use of the co-clustering technique for constructing the codebook 
from the source domain, i.e., for obtaining the cluster-level rating pattern. After this stage, codebook is processed by comparing the entries of codebook with the values of the filled-in source rating matrix. 
By applying the maximum margin matrix factorization technique on the processed codebook, the latent factor vectors of codebook are obtained. The learnt source knowledge (latent factors of codebook) is then transferred to the target domain via hinge loss, and the target domain latent features are learned which are then utilized to get the predicted target user-item rating matrix. By observing the experimental results on the benchmark data sets, we say that our model approximates the target matrix well. In the proposed method the only consideration we gave is for the ratings and it is possible in the future to consider \emph{social tags} also as input types. Transfer learning applications of other type rather than rating prediction in recommender systems could be another direction that could be pursued. 

\section*{Acknowledgements}
This work has been done as part of the PhD dissertation of the first
author at University of Hyderabad. The first author would like
to acknowledge the funding agency, Council of Scientific and
Industrial Research (CSIR) Government of India for the financial
support in the form of CSIR-UGC NET-JRF/SRF.

\bibliographystyle{apalike}
\bibliography{mybibfile}

\end{document}